# PHASE TRANSITION IN THE SIMPLEST PLASMA MODEL

I.L. Iosilevski

*Moscow Institute of Physics and Technology, Dolgopudny, Moscow region, Russia*

We have investigated the phase transition of the gas-liquid type, with an upper critical point, in a variant of the One Component Plasma model (OCP) that has a uniform but compressible compensating background. We have calculated the parameters of the critical and triple points, spinodals, and two-phase coexistence curves (binodals). We have analyzed the connection of this simplest plasma phase transition with anomalies in the spatial charge profiles of equilibrium non-uniform plasma in the local-density approximations of Thomas-Fermi or Poisson-Boltzmann-type.

**1. Introduction**

The simplest plasma model is the One-Component Plasma (OCP), which is a system of free, mobile charges of the same sign in a uniform, compensating background of opposite sign [1]. In the ordinary version of the model - the OCP with a rigid background - volume variations are not defined. The only phase transition - crystallization - occurs without any change of density.

The main subject of our present interest is the OCP with a compressible but still uniform background. This means that the background does not screen the moving charges individually but does so only on average. This variant of the OCP naturally follows from the variational principle of statistical mechanics when we first "switch off" the correlations between positive and negative charges and after that the correlations inside one of the subsystems. It is known that crystallization occurs with a small density variation in this model [2]. The main statement of the present report is the appearance of a new first-order phase transition [3], the properties of which strongly depend on the precise definition of the background thermodynamics.

**2. Definition of the model**.

Several different versions of the OCP in a compressible background can be distinguished:

**A**) An OCP with variable density of the uniform background - which may obey a purely electrostatic, or ideal gas, or any other equation of state - can be obtained through the overall electroneutral grand canonical ensemble4.

**B**) The background thermodynamics is described only through a Helmholtz free-energy-density functional with a strong gradient correction. This suppresses the background non-uniformities, so that the strength tends to infinity after the thermodynamic limit ($N \to \infty$, $N/V = const$).

**C**) We may obtain an overall density-functional description with the pair correlation functions as independent arguments, setting

$$n_{ei}(\mathbf{x},\mathbf{y}) \equiv n_e(\mathbf{x})n_i(\mathbf{y}); \qquad n_{aa}(\mathbf{x},\mathbf{y}) \equiv n_a(\mathbf{x})n_a(\mathbf{y}); \qquad (\text{a = either electrins or ions}). \qquad (1)$$

The main consequence of these different definitions of the OCP is the additivity of the resulting equation of state. Thus, when we know the equations of state of both subsystems (OCP and background), we can establish the existence of the phase transition and carry out direct calculations of its parameters.

### 3. Classical Point Charges in an Ideal Gas Background.

When the background is a Boltzmann ideal gas, the OCP is thermodynamically unstable against collapse at any temperature and density. When the background obeys Fermi statistics, a phase transition of the gas-liquid type, with an upper critical point, appears at a sufficiently low temperature or at sufficiently large values of the charge number. The results of our calculations of the phase transition parameters defined in Eq. (2) are presented in Table I in standard notation.

$$\Gamma \equiv \frac{Ze^2}{kT}\left(\frac{4\pi n_Z}{3}\right)^{1/3} \qquad \Lambda_e^2 \equiv \frac{2\pi\hbar^2}{m_e kT} \qquad r_s \equiv a_B^{-1}\left(\frac{4\pi n_e}{3}\right)^{-1/3} \qquad a_B \equiv \frac{\hbar^2}{m_e e^2} \qquad (2)$$

In these calculations, we used analytical fits from Ref. 1 (for the OCP) and from Ref. 5 (for the ideal Fermi gas).

| | Z | 1 | 2 | 3 | 10 | 30 | 100 |
|---|---|---|---|---|---|---|---|
| I | $T_c$, Ry | 0.0397 | 0.151 | 0.311 | 2.13 | 10.3 | 54.0 |
| | $\Gamma_e$ | 9.09 | 13.1 | 17.1 | 44.5 | 122 | 393 |
| | $(r_s)_c$ | 5.54 | 3.21 | 2.35 | 9.81 | 0.459 | 0.203 |
| | $(n_e\Lambda^3)_c$ | 7.89 | 5.45 | 4.71 | 3.64 | 3.30 | 3.20 |
| | $\{P/(n_z + n_e)kT\}_c$ | 0.126 | 0.126 | 0.125 | 0.124 | 0.123 | 0.121 |
| II | $T_{tr}$ [Ry] | 0.0049 | 0.0246 | 0.0633 | 1.02 | 10.0 | 39.5 |
| | $(\Delta n/n)_{tr}$ | 0.019 | 0.019 | 0.019 | 0.022 | 0.13 | 0.007 |
| III | $(r_s)_{bin}$ | 2.47 | 1.55 | 1.19 | 0.53 | 0.255 | 0.115 |
| | $(r_s)_{spin}$ | 3.08 | 1.94 | 1.48 | 0.66 | 0.319 | 0.143 |
| | $\Gamma_{spin}$ | 5.76 | 8.39 | 11.0 | 28.9 | 79.6 | 256 |
| | $(\Delta H_f/N_z)_0$, Ry | 0.3633 | 1.831 | 4.715 | 78.2 | 1016 | 16860 |

Table I. The phase transition parameters for an OCP consisting of classical point charges in an ideal Fermi-gas background, for different values of the charge number Z: I – critical point; II – triple point; III – the parameters of condensed ($r_s$) and gaseous ($\Gamma_{spin}$) binodals and spinodals, and the heat of sublimation ($\Delta H_f/N_Z$) per ion in the limit $T \to 0$.

We can make some comments about these results:

1) The specially constructed compressibility factor $\{P/(n_z + n_e)kT\}_c$ at the critical point has almost the same (rather low) value for all Z.

2) When the triple point does not coincide with the critical point ($Z \neq Z^* \approx 44$) the density variation between crystalline and fluid phases is fairly small (see $(\Delta n/n)_{tr}$ Tab. I) although much greater than estimated earlier [2] for the case $r_s \ll 1$ ($\Delta V/V \approx 0.0003$). The density variation increases remarkably when $Z \to Z^*$.

3) For $Z > Z^*$ the melting curve ($\Gamma \cong 160$) crosses the "gaseous" part of the binodal and forms the specific picture of a phase diagram.

4) The deviation from the well-known semi-empirical rule of the "rectilinear diameter" is similar to those obtained for alkali metals, but the slope of this diameter and the relation of the normal density to the critical density is rather large ($n_0/n_c \cong 12.5$) in comparison with those for real substances. For example, for alkali metals we have $n_0/n_c \cong 4 - 5$.

5) The critical point of this phase transition seems not to be a genuine one because spatial density fluctuations are absent in its vicinity in the model definition "**A**" they are suppressed by the gradient term in definition "**B**" and they are only partially admitted in definition "**C**". In this latter case (and in the case of a finite gradient term), the position of the critical point must deviate slightly from the calculated limiting value.

6) A remarkable feature of any non-uniformity in the equilibrium system of Coulomb particles (in particular, of the inter-phase boundary) is a finite difference in the average electrostatic potential across this non-uniformity. The equality of two electrochemical potentials for every kind of Coulomb particles on both sides of the non-uniformity leads to an inter-phase potential drop depending only on the temperature. Note the high value of this drop for the OCP model at $T \to 0$: $e\Delta\varphi/kT_c = 19.5$. Estimates for some metals give $e\Delta\varphi/kT_c \approx 2$ to $3$.

**4. Quantum Electron Gas**

In the variant of the OCP with a rigid background, the only phase transition is still crystallization, with only the change that "cold melting" appears in the high-density limit [6,7]. The phase diagram changes in an essential way when we deal with a uniform but compressible background because of the appearance of a vast phase transition of the gas-liquid type. The parameters of this phase transition, calculated from analytical fits, are:

$$T_c = 0.0478 \text{ Ry}; \quad (r_s)_c = 9.27; \quad \Gamma_c = 4.53; \quad (n_e \Lambda_e^3)_c = 1.28. \quad (3)$$

These results are close to those obtained previously (Table I). Thus, we may conclude that this new phase transition completely excludes Wigner crystallization, because its melting curve [6,7] is *located deep inside* the gas-liquid spinodal curve. At all temperature this crystal is thus absolutely unstable *against a phase* decomposition into two weakly coupled fluid plasmas - a *dense* and a *rare* one. This was previously suggested for $T \approx 0$.

**5. The Double-OCP Model**

When using the variational principle of statistical mechanics, we may "switch off" only correlations between charges of opposite sign. In this case, the nuclei and electrons become compressible, compensating backgrounds for each other. This combination of two OCP models, inserted one into the other, will be referred to as a "Double OCP." By "switching off" other kinds of dynamic correlations inside each subsystem we can obtain a hierarchy of the simplest plasma models with decreasing free energies [3, 10]. When we assume the whole system to be uniform, the resulting equation of state is the sum of those for the two separate subsystems. The phase diagram of the Double OCP depends on the masses of the charges. For heavy particles, it is similar to the one described in §3. The main difference is "cold melting" of the nuclear subsystem in the high-density limit. The parameters of the critical point for this case are also quite close to those of Table I:

$$T_c = 0.1008 \text{ Ry}; \qquad (r_s)_c = 3.58; \qquad \Gamma_c = 5.54; \qquad (n_e \Lambda_e^3)_c = 0.724. \qquad (4)$$

Note also the decrease of the critical compressibility factor $\{P/(n_z + n_e)kT\}_c \approx 0.1074$ (!). For mass-symmetrical plasmas (e.g., electron-positron and electron-hole plasmas) the phase diagram coincides with that of the electron gas. Thus, only a phase transition of the gas-liquid type exists. The Double-OCP model gives a simple estimate of the critical temperature of this phase transition. For example, for an electron-positron plasma it is $kT_c \approx 0.65$ eV.

**6. OCP Phase transition and anomalies in the equilibrium spatial charge distribution**

Let us consider situations where our knowledge of the properties of this phase transition may be useful. This is the case, for example, in any spatial charge profile calculations in equilibrium, non-uniform plasmas. To be more precise, this is the case when we use the local equation of state approximation in the simplest density-functional description [11] separately for each kind of charge and try to improve the well-known ideal-gas approximations of Thomas-Fermi or Poisson-Boltzmann using the local exchange and correlation corrections. When we apply the exact OCP free-energy density, or the simple Dirac- or Debye-type corrections, for this purpose, the discussed macroscopic phenomenon - the phase transition - must display itself in the results of calculations as a discontinuity in the spatial charge profile at sufficiently low temperatures and densities. In terms of the OCP phase transition this boundary temperature is just the critical temperature, and the two values of the local densities at the discontinuity are the coexisting condensed and gaseous densities, which depend only on the temperature. For example, in an atomic cell the electron profile calculated for $T < T_c$ and $n < n_e$ with this approximation must break up into a condensed "drop" around an attractive center and diffuse "atmosphere" at the cell periphery. From our point of view, this is precisely the phenomenon that well-known as a finite-size atom in the Thomas-Fermi-Dirac approximation; it may be seen in figures 12 to 15 in Gombas's book [12]. Note that in the $T = 0$ case the density of the "gaseous" part of the electron profile equal to zero. For $0 \neq T < T_c \approx 0.65$ eV, and for the same range of density, the discussed TFD-electron gas profile discontinuity must lead to a discontinuity in the "degree of ionization" in the atomic cell. This was calculated [13] using the cell boundary electron density. Unfortunately the results for these temperatures re omitted in Ref. 13. The supposed structure of this discontinuity at $T \leq T_c$ is shown schematically in Ref. 10.

**Acknowledgements**

The author is grateful to V. Fortov and E. Son for their friendly encouragement and interest in this work and to A. Chigvintsev for great help in numerical calculations. The author thanks also the Organizing Committee and the MGMP-company "Agroremmonitor" for their support which helped the author to attend the conference.